\documentclass[12pt]{article}
\usepackage{putex}
\begin{document}

\preprint{hep-th/0605182 \\ PUPT-2198}

\institution{PU}{Joseph Henry Laboratories, Princeton University, Princeton, NJ 08544}

\title{Drag force in AdS/CFT}

\authors{Steven S. Gubser}

\abstract{The AdS/CFT correspondence and a classical test string approximation are used to calculate the drag force on an external quark moving in a thermal plasma of ${\cal N}=4$ super-Yang-Mills theory.  This computation is motivated by the phenomenon of jet-quenching in relativistic heavy ion collisions.}

\date{May 2006}

\maketitle

\section{Introduction}

A calculation of the ratio of shear viscosity to entropy density of near-extremal D3-branes \cite{Policastro:2001yc} raised the tantalizing prospect of a connection between string theory and relativistic heavy ion collisions.  The connection hinges on two premises:
 \begin{enumerate}
  \item Experiments at RHIC \cite{Arsene:2004fa,Adcox:2004mh,Back:2004je,Adams:2005dq} probe properties of the quark gluon plasma (QGP) a thermal state of matter where colored states are deconfined, but the coupling is still strong: the 't~Hooft coupling is on the order of $10$.
  \item Near-extremal D3-branes provide a gravitational representation of ${\cal N}=4$ super-Yang-Mills theory at finite temperature \cite{Gubser:1996de}, large $N$, and strong 't~Hooft coupling, $g_{YM}^2 N \gg 1$.
 \end{enumerate}
In most circumstances, replacing QCD by ${\cal N}=4$ super-Yang-Mills can be charitably described as an uncontrolled approximation.  The gauge fields and their tree-level interactions are identical between the two theories, but the matter fields and the quantum dynamics are different in many ways.  A conservative view, then, is that near-extremal D3-branes are merely an analogous system to the QGP.  A more optimistic view is that ${\cal N}=4$ super-Yang-Mills at a given $g_{YM}^2 N \gg 1$ and finite temperature $T$ may capture much of the dynamics of the QGP in a range of temperatures high enough so that confinement and the chiral condensate have disappeared but low enough so that the 't~Hooft coupling is still large.  In view of the success of \cite{Policastro:2001yc} in at least qualitatively reproducing the low shear viscosity that events with elliptic flow appear to exhibit, as well as the difficulty of formulating any other analytic framework which incorporates strong-coupling quantum dynamics, one can't help hoping that more can be learned about heavy ion physics from the near-extremal D3-brane and its variants.  Indeed, a significant literature has emerged on this topic, with which my familiarity is regrettably slight: see for example \cite{Policastro:2002se,Shuryak:2004cy,Nastase:2005rp,Shuryak:2005ia} and references therein.

One of the more distinctive features of RHIC data is the phenomenon of jet-quenching: strong energy loss as a high-energy parton passes through the QGP.  To make a connection with string theory, one should start by passing some sort of colored object through the finite temperature ${\cal N}=4$ super-Yang-Mills plasma.  Readiest to hand is an external quark, which in the framework of AdS/CFT \cite{Maldacena:1997re,Gubser:1998bc,Witten:1998qj} is represented as a string dangling from the boundary of $AdS_5$-Schwarzschild \cite{Maldacena:1998im,Rey:1998ik,Rey:1998bq,Brandhuber:1998bs,Danielsson:1998wt}.  Its point of attachment on the boundary carries a fundamental charge under the gauge group $SU(N)$, and it is infinitely massive.  

The external quark can be prescribed to move on the boundary of $AdS_5$-Schwarzschild along any timelike trajectory.  Because the quark is infinitely massive, the most sensible trajectories are those with constant velocity relative to the reference frame defined by the timelike Killing vector of $AdS_5$-Schwarzschild (that is, the rest frame of the plasma).  One then expects that the string trails out behind the quark, arcing down into $AdS_5$-Schwarzschild.  The string is a holographic representation of the color flux from the external quark spreading out in the $3+1$ dimensions of the boundary theory.  The string exerts a drag force on the external quark, the calculation of which is the main aim of this paper.  The magic of AdS/CFT is that one can use a classical picture of a string in curved space to capture strong coupling gauge dynamics.  But we must keep in mind that the underlying theory is ${\cal N}=4$ super-Yang-Mills (or some other theory known to have an $AdS_5$ dual) rather than real-world QCD.

While this paper was in preparation, I received \cite{Herzog:2006}, which includes the main results of this work and extends them in several ways, and also \cite{Liu:2006ug}, in which a different approach is taken to describing jet-quenching in AdS/CFT.

\section{Test strings in $AdS_5$-Schwarzschild}

The metric of the near-extremal D3-brane is
 \eqn{NearDThree}{
  ds_{10}^2 &= H^{-1/2} (-h dt^2 + d\vec{x}^2) + 
   H^{1/2} \left( {dr^2 \over h} + d\Omega_5^2 \right)  \cr
  H &= 1 + {L^4 \over r^4} \qquad 
  h = 1 - {r_H^4 \over r^4} \,.
 }
Here $\vec{x} = (x,y,z)$ are the spatial coordinates along which the D3-brane is extended and $d\Omega_5^2$ is the standard metric on the five-sphere $S^5$ with unit radius.  The near-horizon limit consists of ``dropping the 1'' from $H$.  Then the metric is $AdS_5$-Schwarzschild,
 \eqn{AdSSchwarzschild}{
  ds_5^2 = G_{\mu\nu} dx^\mu dx^\nu = 
    {r^2 \over L^2} (-h dt^2 + d\vec{x}^2) + 
    {L^2 \over r^2} {dr^2 \over h} \,,
 }
times the metric for an $S^5$ of constant radius $L$.  The crucial insight of \cite{Maldacena:1997re} is that this asymptotically anti-de Sitter part of the geometry is the part that encodes the physics of ${\cal N}=4$ super-Yang-Mills theory.

A test string in $AdS_5$-Schwarzschild can be described by the Nambu-Goto action:
 \eqn{NGaction}{
  S = -{1 \over 2\pi\alpha'} \int d^2 \sigma \, e^{\phi/2}
    \sqrt{-\det g_{\alpha\beta}} \qquad
  g_{\alpha\beta} \equiv G_{\mu\nu} \partial_\alpha X^\mu
    \partial_\beta X^\nu \,,
 }
where $\sigma^\alpha$ are coordinates on the string worldsheet, $X^\mu = X^\mu(\sigma)$ specifies the embedding of the string worldsheet in spacetime, and $G_{\mu\nu}$ is the $AdS_5$-Schwarzschild metric \eno{AdSSchwarzschild}.  The dilaton factor $e^{\phi/2}$ accounts for the fact that $G_{\mu\nu}$ is the five-dimensional Einstein metric.  This factor can be dropped because $\phi=0$ in the background.  The equations of motion following from \eno{NGaction} may be expressed as
 \eqn{NGeoms}{
  \nabla_\alpha P^\alpha{}_\mu = 0 \qquad
   P^\alpha{}_\mu \equiv -{1 \over 2\pi\alpha'} 
     G_{\mu\nu} \partial^\alpha X^\nu \,,
 }
where $\nabla_\alpha$ is the covariant derivative with respect to $g_{\alpha\beta}$.  $P^\alpha{}_\mu$ is the worldsheet current of spacetime energy-momentum carried by the string \cite{Lawrence:1993sg}.

Consider motions of the string in only one of the spatial directions spanned by the D3-brane: the $x^1$-direction.  In static gauge, $\sigma^\alpha = (t,r)$, the embedding of the worldsheet is completely specified by the function $x(t,r)$.  Without assuming that this function is small, one may obtain from \eno{NGaction}
 \eqn{NGspecial}{
  S = {1 \over 2\pi\alpha'} \int dt dr \, {\cal L} \qquad
  {\cal L} = -\sqrt{1 + {h \over H} x'^2 - {\dot{x}^2 \over h}} \,,
 }
where $\dot{x}$ means $\partial x / \partial t$ and $x'$ means $\partial x / \partial r$.  For brevity I have written $H$ in place of $L^4/r^4$.

\section{Drag force on an external quark}

A suitable ansatz to describe the late-time behavior of a string attached to an external quark that moves with speed $v$ in the $x^1$ direction is (in static gauge)
 \eqn{LateTimes}{
  x^1(t,r) = v t + \xi(r) + o(t) \,,
 }
where the $o(t)$ term includes all other motions which damp out at late times.  The ansatz \eno{LateTimes} depends on the assumption that steady state behavior is achieved at late times.  From now on we will neglect the $o(t)$ part entirely.

It is easy to show that the differential equation one gets for $\xi(r)$ is the same if one plugs \eno{LateTimes} into the Euler-Lagrange equations following from \eno{NGspecial} or if one first substitutes \eno{LateTimes} into the lagrangian \eno{NGspecial} and then derives the equations of motion.  The latter approach is a little more transparent: one has
 \eqn{Lred}{
  L &= -\sqrt{1 - {v^2 \over h} + {h \over H} \xi'^2} \qquad
  \pi_\xi \equiv {\partial L \over \partial \xi'} \,,
 }
and the equation of motion is simply that $\pi_\xi$ is a constant.  One may solve the defining relation for $\pi_\xi$ for $\xi'$:
 \eqn{FoundYp}{
  \xi' = \pm\pi_\xi {H \over h} \sqrt{h-v^2 \over h - \pi_\xi^2 H} \,.
 }
To fix the sign ambiguity, observe that $\xi'$ should be positive in order to describe a string that trails out behind the external quark rather than ``trailing'' in front of it.  If $\pi_\xi$ is chosen positive, then the sign in \eno{FoundYp} should be $+$.

There is an obstacle to integrating \eno{FoundYp} to find $\xi(r)$: we must require that $\xi(r)$ is everywhere real, but the square root on the right hand side is in general not everywhere real.  The function $h(r)$ interpolates between $1$ at the boundary of $AdS_5$-Schwarzschild to $0$ at the horizon, so at some intermediate radius $h-v^2$ switches sign.  The only way to avoid an imaginary right hand side in \eno{FoundYp} is to set 
 \eqn{SpecialPiXi}{
  \pi_\xi = {v \over \sqrt{1-v^2}} {r_H^2 \over L^2} \,.
 }
It is interesting that $\pi_\xi$ is set not by a boundary condition exactly at the horizon, but rather at some larger radius.  Plugging \eno{SpecialPiXi} into \eno{FoundYp} leads to a remarkably simple result:
 \eqn{SimpleXi}{
  \xi' &= v {r_H^2 \over L^2} {H \over h} = 
    v {r_H^2 L^2 \over r^4 - r_H^4}  \cr
  \xi &= -{L^2 \over 2r_H} v \left( \tan^{-1} {r \over r_H} + 
    \log\sqrt{r+r_H \over r-r_H} \right) \,.
 }

We have already remarked that $P^\alpha{}_\mu$ defined in \eno{NGeoms} is the conserved worldsheet current of spacetime energy-momentum.  To calculate the flow of momentum $dp_1/dt$ down the string (that is, from the external quark into infrared dissipation) one needs the following integral:
 \eqn{DeltaP}{
  \Delta P_1 = 
    \int_{\cal I} dt \, \sqrt{-g} P^r{}_{x^1} = {dp_1 \over dt}
     \Delta t \,.
 }
where the integral is taken over some time interval ${\cal I}$ of length $\Delta t$.  Because $P^\alpha{}_\mu$ is conserved, it should not matter at what radius we choose to evaluate the integral.
The orientation of the integral is chosen so that $dp_1/dt$ is a negative quantity: it is the drag force, which points opposite the motion.  A slightly tedious calculation leads to
 \eqn{dpdt}{
  {dp_1 \over dt} = \sqrt{-g} P^r{}_{x^1} = 
    -{\sqrt{1-v^2} \over 2\pi\alpha'} G_{x^1\nu} g^{r\alpha} 
      \partial_\alpha X^\nu = 
    -{r_H^2/L^2 \over 2\pi\alpha'} {v \over \sqrt{1-v^2}} \,.
 }
Now we recall two useful formulas:\footnote{See for example \cite{Klebanov:2000me}.  But note that modern usage is $g_{YM}^2 = 4\pi g_{\rm string}$ rather than $g_{YM}^2 = 2\pi g_{\rm string}$, leading to factors as written in \eno{dimensionless}.}
 \eqn{dimensionless}{
  L^4 = g_{YM}^2 N \alpha'^2  \qquad  T = {r_H \over \pi L^2} \,,
 }
where $T$ is the Hawking temperature, whose dual description is nothing but the temperature of the plasma.  Plugging \eno{dimensionless} into \eno{dpdt} leads to the final result for the drag force:
 \eqn{DragForceOne}{
  {dp_1 \over dt} = -{\pi \sqrt{g_{YM}^2 N} \over 2} 
    T^2 {v \over \sqrt{1-v^2}} \,.
 }

It is interesting to express the result \eno{DragForceOne} in terms of the momentum $p_1$ and mass $m$ of the external quark, even though both are formally infinite:
 \eqn{DragForceTwo}{
  {dp_1 \over dt} = -{\pi \sqrt{g_{YM}^2 N} \over 2} T^2 
    {p_1 \over m} \,.
 }
Continuing in this somewhat formal spirit, one can integrate \eno{DragForceTwo} to find
 \eqn{pVSt}{
  p_1(t) = p_1(0) e^{-t/t_0} \qquad
    t_0 = {2 \over \pi \sqrt{g_{YM}^2 N}} {m \over T^2} \,.
 }
Keeping previous remarks about uncontrolled approximations firmly in mind, consider the following equivalent ways of expressing $t_0$ in physical units:
 \eqn{CharmBottom}{
  \hbox{bottom:} &\qquad
   t_0 \approx 2\,{\rm fm}/c \ {m/m_b \over \sqrt{g_{YM}^2 N/10} \;
     (T / 300\,{\rm MeV})^2}
    \cr
  \hbox{charm:} &\qquad
   t_0 \approx 0.6\,{\rm fm}/c \ {m/m_c \over \sqrt{g_{YM}^2 N/10} \;
     (T / 300\,{\rm MeV})^2} \,.
 }
It might be illuminating to compare these relaxation times with the rate of energy loss of bottom and charm quarks moving through the QGP.

\section{Discussion}

The expressions \eno{CharmBottom} for relaxation times are at least in an interesting ballpark for RHIC experiments, where timescales of interest are between roughly $0.3\,{\rm fm}/c$ and $8\,{\rm fm}/c$.  But the mechanism is very different from more established theoretical descriptions of jet-quenching at RHIC, for example the framework (\cite{Gyulassy:2000fs,Wang:2001cs,Baier:2001yt,Salgado:2003gb} and references therein) in which energy loss depends on the square of the distance traveled because of destructive interference effects between the quark and a radiated gluon.

Carrying the formulas \eno{DragForceTwo} and \eno{pVSt} over to RHIC physics makes the most sense for heavy quarks, as we have suggested in \eno{CharmBottom}.  But does it make sense even there?  The absence of dynamical quarks in the fundamental representation in ${\cal N}=4$ super-Yang-Mills means that the QCD string cannot break, whereas of course it does in QCD.  Perhaps the thermal mass for quarks inhibits this enough for the trailing string in $AdS_5$-Schwarzschild to have some validity.  This trailing string has a dual description, perhaps best thought of as the shadow it casts on the boundary, which could be computed by finding the dilaton and metric perturbations generated by the string and propagating them out to the boundary, as in \cite{Danielsson:1998wt,Callan:1999ki}.  The result would be profiles for $\langle \tr F^2 \rangle$ and $\langle T_{\mu\nu} \rangle$ in the gauge theory, which for highly relativistic velocities presumably indicate a sort of ``wake'' with the external quark at its leading tip.\footnote{In fact, the dilaton couples to an operator ${\cal O}_{F^2}$ which is (up to a subtlety regarding on-shell terms) the lagrangian density of ${\cal N}=4$ super-Yang-Mills.  So it is more precise to say that the dilaton ``shadow'' cast by the string on the boundary translates into a VEV $\langle {\cal O}_{F^2} \rangle$.}  Perhaps the shape of such profiles could be related to the topology of individual events where a hard jet dumps most of its energy into the QGP.

\section*{Acknowledgments}

I thank B.~Zajc for suggesting that I look into a string theory analog of jet quenching and for commenting on a draft of the paper.  This work was supported in part by the Department of Energy under Grant No.\ DE-FG02-91ER40671, and by the Sloan Foundation.

\bibliographystyle{ssg}
\bibliography{quark}

\end{document}